\documentclass[reprint, superscriptaddress, preprintnumbers, amsmath, amssymb, aps, floatfix]{revtex4-2}

\usepackage{graphicx}
\usepackage{dcolumn}
\usepackage{bm}
\usepackage[utf8]{inputenc}
\usepackage[T1]{fontenc}
\usepackage{mathptmx}
\usepackage{textcomp}
\usepackage{siunitx}
\usepackage{bigints}
\usepackage{subfig}
\usepackage[colorlinks]{hyperref}
\hypersetup{allcolors = {black}}
\usepackage[switch]{lineno}
\usepackage[table,xcdraw,dvipsnames]{xcolor}
\usepackage{booktabs}
\usepackage{float}
\usepackage{color}
\usepackage{multirow}
\makeatletter
\AtBeginDocument{%
  \renewcommand\normalsize{\@setfontsize\normalsize{11pt}{13.2pt}}%
  \normalsize
}
\makeatother

\usepackage{caption}
\DeclareCaptionFont{tenpt}{\fontsize{10pt}{12pt}\selectfont}
\hypersetup{colorlinks=true,allcolors=blue,breaklinks=true}

\usepackage{titlesec}

\makeatletter
\renewcommand\thesection{\arabic{section}}
\renewcommand\thesubsection{\thesection.\arabic{subsection}}
\renewcommand\thesubsubsection{\thesubsection.\arabic{subsubsection}}
\makeatother

\titleformat{\section}
  {\normalfont\bfseries\large}{\thesection.}{0.6em}{}
\titleformat{\subsection}
  {\normalfont\bfseries\normalsize}{\thesubsection.}{0.6em}{}
\titleformat{\subsubsection}
  {\normalfont\bfseries\normalsize}{\thesubsubsection.}{0.6em}{}

\titlespacing*{\section}{0pt}{*3}{*1}
\titlespacing*{\subsection}{0pt}{*2}{*0.7}
\titlespacing*{\subsubsection}{0pt}{*1.5}{*0.5}

\makeatletter
\renewcommand{\selectlanguage}[1]{}
\makeatother

\begin{document}
\title[AEPSWS]{k-Selective Electrical-to-Magnon Transduction with Realistic Field-distributed Nanoantennas}

\author{Andreas~H\"ofinger}
\email{andreas.hoefinger@univie.ac.at}
\affiliation{Nanomagnetism and Magnonics, Faculty of Physics, University of Vienna, A-1090 Vienna, Austria}%

\author{Andrey~A.~Voronov}
\affiliation{Nanomagnetism and Magnonics, Faculty of Physics, University of Vienna, A-1090 Vienna, Austria}%
\affiliation{Vienna Doctoral School in Physics, University of Vienna, A-1090 Vienna, Austria}%

\author{David~Schmoll}
\affiliation{Nanomagnetism and Magnonics, Faculty of Physics, University of Vienna, A-1090 Vienna, Austria}%
\affiliation{Vienna Doctoral School in Physics, University of Vienna, A-1090 Vienna, Austria}%

\author{Sabri~Koraltan}
\affiliation{Institute of Applied Physics, TU Wien, Wiedner Hauptstra{\ss}e 8-10, A-1040 Vienna, Austria}%
\affiliation{Physics of Functional Materials, Faculty of Physics, University of Vienna, A-1090 Vienna, Austria}%

\author{Florian~Bruckner}
\affiliation{Physics of Functional Materials, Faculty of Physics, University of Vienna, A-1090 Vienna, Austria}%

\author{Claas~Abert}
\affiliation{Physics of Functional Materials, Faculty of Physics, University of Vienna, A-1090 Vienna, Austria}%
\affiliation{Research Platform MMM Mathematics-Magnetism-Materials, University of Vienna, A-1090 Vienna, Austria}%

\author{Dieter~Suess}
\affiliation{Physics of Functional Materials, Faculty of Physics, University of Vienna, A-1090 Vienna, Austria}%
\affiliation{Research Platform MMM Mathematics-Magnetism-Materials, University of Vienna, A-1090 Vienna, Austria}

\author{Morris~Lindner}
\affiliation{INNOVENT e.V. Technologieentwicklung, Prüssingstraße 27B, 07745 Jena, Germany}

\author{Timmy~Reimann}
\affiliation{INNOVENT e.V. Technologieentwicklung, Prüssingstraße 27B, 07745 Jena, Germany}

\author{Carsten~Dubs}
\affiliation{INNOVENT e.V. Technologieentwicklung, Prüssingstraße 27B, 07745 Jena, Germany}

\author{Andrii~V.~Chumak}
\email{andrii.chumak@univie.ac.at}
\affiliation{Nanomagnetism and Magnonics, Faculty of Physics, University of Vienna, A-1090 Vienna, Austria}%

\author{Sebastian~Knauer}
\email{knauer.seb@gmail.com}
\affiliation{Nanomagnetism and Magnonics, Faculty of Physics, University of Vienna, A-1090 Vienna, Austria}%


\begin{abstract}
The excitation and detection of propagating spin waves with lithographed nanoantennas underpin both classical magnonic circuits and emerging quantum technologies.
Here, we establish a framework for all-electrical propagating spin-wave spectroscopy (AEPSWS) that links realistic electromagnetic drive fields to micromagnetic dynamics. 
Using finite-element~(FE) simulations, we compute the full vector near-field of electrical impedance-matched, tapered coplanar and stripline antennas and import this distribution into finite-difference~(FD) micromagnetic solvers. 
This approach captures the antenna-limited wave-vector spectrum and the component-selective driving fields (perpendicular to the static magnetisation) that simplified uniform-field models cannot. 
From this coupling, we derive how realistic current return paths and tapering shapes, k-weighting functions, for Damon–Eshbach surface spin waves in yttrium-iron-garnet (YIG) films are, for millimetre-scale matched CPWs and linear tapers down to nanometre-scale antennas.
Validation against experimental AEPSWS on a 48\,nm YIG film shows quantitative agreement in dispersion ridges, group velocities, and spectral peak positions, establishing that the antenna acts as a tunable k-space filter.
These results provide actionable design rules for on-chip magnonic transducers, with immediate relevance for low-power operation regimes and prospective applications in quantum magnonics. 
\end{abstract}
\maketitle

\textbf{Keywords:} All-electrical propagating spin-wave spectroscopy, Nanoantenna magnonic transducers, Coupled finite-element–micromagnetic modelling
\vspace{0.3cm}

\section{\label{sec:level0}Introduction}
All-Electrically driven propagating spin-wave spectroscopy (AEPSWS) excites and detects spin waves, and their quasi-particles, magnons, using on-chip microwave transducers patterned on low-loss magnetic films such as yttrium-iron-garnet (YIG) on gadolinium-gallium-garnet (GGG)~\cite{Serga2010, Dubs2017, Liu2018, Dubs2020, Will-Cole2023}. In modelling practice, antennas are often represented by uniform or single-component RF fields to estimate excitation spectra and dispersions, while micromagnetic solvers propagate the magnetisation dynamics~\cite{Bance2008, Durflou2017, Copus2022, Adrien2024}. This approach is practical and widely used, but it abstracts away the geometry-, taper-, and impedance-dependent near-field vector distribution of realistic on-chip transducers and their \emph{k}-selective coupling to spin-wave modes, often to Damon–Eshbach modes. Related electromagnetic (EM) models have optimised antenna impedance and current distribution~\cite{RAO2019, ZHANG2018, Mori_2021}, typically without resolving the resulting \emph{k}-selective coupling to spin waves.

In modern devices, the nanoantenna acts as a hard spectral filter: conductor widths, taper profiles, and current return paths sculpt both the accessible wave-vector content and the mix of field components that reach the magnetic film~\cite{Ciubotaru2016, Heinz2020, Knauer2023, Vanatka2021}. Only the component of the RF field perpendicular to the static magnetisation,
$\mathbf h_{\!\perp}=\mathbf h-(\mathbf h\!\cdot\!\hat{\mathbf M}_0)\hat{\mathbf M}_0$, efficiently drives spin precession, neglecting the full 3D near-field (including its phase) prevents one from (i) predicting which $\mathbf k$'s are actually launched, (ii) quantifying electrical-to-magnon transduction as a function of frequency, and (iii) optimising tapers and line impedances for targeted, \emph{k}-selective spectra. These limitations matter not only for classical magnonic circuits~\cite{Chumak2015, Kruglyak2010, Barman2021, Casulleras2023, Wang2024, Flebus2024} but also for low-excitation regimes relevant to quantum magnonics~\cite{Demokritov2006, Lachance-Quirion2019, Lachance-Quirion2020, Pirro2021, YUAN2022, Chumak2022, Xu2023, Kostylev2023, Dobrovolskiy2025}.

Here, we couple a frequency-domain EM finite-element (FE) model of a $50\,\Omega$-matched, linearly tapered coplanar waveguide (CPW) signal line and CPW/stripline nanoantennas (for RF drive), to a micromagnetic finite-difference (FD) solver by importing the complex 3D vector near-field at the YIG surface as the drive. In $(\mathbf k,\omega)$-space we compute
\(
 \mathbf m(\mathbf k,\omega)=\boldsymbol\chi(\mathbf k,\omega)\,\mathbf h_{\!\perp}(\mathbf k,\omega),   
\)
which yields the antenna \emph{k}-weighting function $W(\mathbf k,\omega)=\|\mathbf h_{\!\perp}(\mathbf k,\omega)\|^2$. Using this end-to-end chain, from millimetre-scale matched CPWs and linear tapers down to nanometre-scale antennas, we quantify how realistic current return paths and tapering shape $W(\mathbf k,\omega)$ for Damon–Eshbach surface spin waves in a $48\,\mathrm{nm}$ YIG film on GGG, and we compare CPW vs.\ stripline excitation antennas fabricated on the same stack.

We focus on the transmission problem (excitation). Receive-side induction in the detector is treated via reciprocity (overlap of the detector’s $\mathbf h_{\!\perp}$ with the propagating $\mathbf m$), explicit EM back-action of the spin wave on the detector conductors is left for future work. We validate simulated dispersions, group velocities, and \emph{k}-peak positions against AEPSWS on lithographed CPW and stripline antennas on the $48\,\mathrm{nm}$ YIG film, using identical stacks and $50\,\Omega$ CPW lines as in the simulations. Experimental context and prior uses of such micro-/nano-antennas include continuous/pulsed excitation~\cite{Vlamninck2008, Haiming2014, Heinz2020, Pirro2013}, integrated circuits~\cite{Wang2020, Talmelli2020, Fischer2017, Heussner2018}, Bose–Einstein condensate control~\cite{Schneider2021}, operation at sub-\textmu{m} scales and mK~temperatures~\cite{Collet2017, Knauer2023}, and engineered transmission bandgaps and exchange-wave excitation~\cite{Li2023, Temdie2023_3, Temdie2023}.

Hybrid EM-spin simulations have been explored with finite-difference time-domain (FDTD) coupling or analytical reductions~\cite{Yao2019, Connelly2021_2, Vanderveken2022}. FDTD excels for strongly time-varying fields, but is computationally intensive for our geometry sweep; FE frequency-domain EM is more efficient for obtaining complex near-fields, when a full time-varying field is not required, which we then inject into FD micromagnetics. For the micromagnetic stage, we use the flexible finite-difference Python package \texttt{magnum.np}~\cite{Bruckner2023}, we compare simulated spectra with the semi-analytical dispersion calculator in \textit{TetraX}~\cite{korberFiniteelementDynamicmatrixApproach2021a, TetraX} and the Kalinikos–Slavin model~\cite{Kalinikos_1986}.

We define the device and antennas in section~\ref{sec:system}, detail the FE and FD methods in section~\ref{sec:methods}, summarise fabrication and measurements in section~\ref{sec:fabrication}, and present results in section~\ref{sec:results}.

\section{Device and antenna geometries}\label{sec:system}
\begin{figure}[tb!]%
    {\includegraphics[width=1.0\columnwidth]{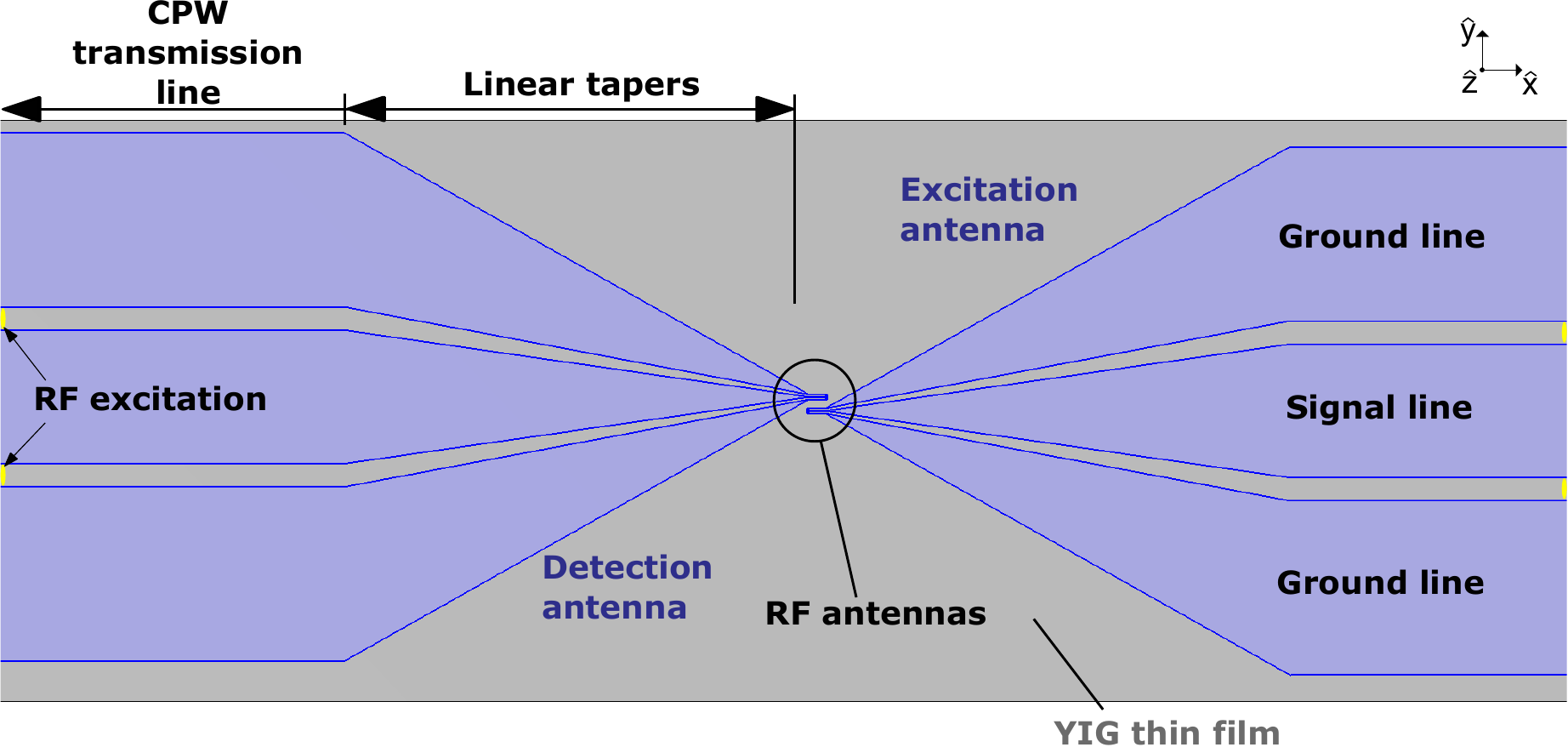}}
    \caption{{\bf The investigated system as an image section of the FE electromagnetic simulation box.} 
    Device stack and RF layout used for AEPSWS. The 48\,nm YIG film (grey) on GGG is contacted by CPW transmission lines with linear tapers to a nanoantenna (stripline or CPW; Sec.~\ref{sec:results})), as shown. Yellow boxes indicate the 50-\(\Omega\) lumped ports used to excite/detect the RF signal, to model a quasi-TEM (Transverse Electromagnetic) transmission line. Coordinate axes are overlaid with \(\hat{\mathbf y}\parallel \mathbf H\) (Damon–Eshbach), \(\hat{\mathbf x}\parallel \mathbf k\), and \(\hat{\mathbf z}\) out of plane. (Colour-coded rendering.) 
    }%
    \label{fig:geometry}
\end{figure}

We consider all-electrical propagating spin-wave spectroscopy~(AEPSWS) on a \qty{48}{\nano\meter}-thin YIG film on a \qty{500}{\micro\meter} GGG substrate. Two on-chip microwave transducers (CPW or stripline) are patterned atop the YIG, the centre-to-centre spacing is \(D=\qty{20}{\micro\meter}\). The in-plane static magnetic field is applied along \(\hat{\mathbf y}\) (Damon–Eshbach), the nominal propagation direction is \(\hat{\mathbf x}\parallel\mathbf k\), and \(\hat{\mathbf z}\) is out of plane. Figure~\ref{fig:geometry} shows the FE simulation snapshot of the system with CPW transmission lines, linear tapers, and the nanoantenna on the YIG/GGG stack; the yellow boxes indicate the \(\qty{50}{\ohm}\) ports used for excitation/detection. The CPW lines used to feed the tapers are designed for \(\sim\qty{50}{\ohm}\) with signal/ground widths of \(\qty{115}{\micro\meter}\), signal–ground spacing \(\qty{20}{\micro\meter}\), and length \(\qty{1}{\milli\meter}\). The FE/FD near-field plots are annotated with the \((x,y,z)\) axes and with the drive-relevant components \(h_x\) and \(h_z\); maps are described as \emph{colour-coded}. The FD micromagnetic box extends \(\qty{70}{\micro\meter}\) along \(x\) so that the spatial envelope of the antenna near-field that seeds propagation is fully included (not only the footprint under the metal). 

\section{Methods}\label{sec:methods}
\subsection{Electromagnetic finite-element model (FE)}

We solve the harmonic RF problem in \textit{COMSOL Multiphysics} (RF module) using \(\qty{50}{\ohm}\) lumped ports at the CPW ends and a linear taper to the nanoantenna, to model a quasi-TEM (transverse electromagnetic) transmission line. The computational domain includes the Ti/Au antenna on YIG/GGG, surrounded by air, and terminated by radiation boundaries/PMLs. For the YIG film, we use room-temperature values for pure YIG: $\epsilon_r=14$, $\mu_r=20$, $\sigma=10^{-8}\,\mathrm{S/cm}$, and for GGG: $\epsilon_r=12$, $\mu_r=1$, $\sigma=10^{-16}\,\mathrm{S/cm}$. Note that, for the proof-of-principle demonstration in this work, we treat our systems as quasi-static, and assume that the permeability and permeability are constant and uniform, to obtain an upper boundary value for the propagation of the quasi-TEM mode in the gigahertz regime. In reality, relative permeability is a complex frequency-dependent tensor. Future work needs to determine how this frequency dependence affects the phase velocity, the effective wavelength, the characteristic impedance, the E/H ratio, and thus the effective magnetic field used for the excitation of spin waves. This degree of freedom would allow the determination of temperature dependences for cryogenic spinwave measurements~\cite{Knauer2023}. One implementation may be the use of a Polder tensor in \textit{COMSOL Multiphysics}. This tensor enables the FE solver to represent gyrotropic, magnetically biased ferrites via an anisotropic, nonreciprocal permeability tensor in Maxwell’s equations. It further captures frequency-dependent resonance and loss in the ferrite, enabling the prediction of nonreciprocal S-parameters and bias-tunable behaviour in full-wave 3D models.

Metals are modelled as lossy conductors; dielectrics are lossless unless noted. Unless stated otherwise, the port power is \(-25~\mathrm{dBm}\). The electromagnetic RF-signal is applied to the CPW transmission line via lumped port boundary conditions, where an alternating voltage of $\qty{12.8}{\milli\volt}$ amplitude with a frequency of $\qty{13}{\giga\hertz}$ is applied (also see Fig.~\ref{fig:geometry}, yellow boxes). 
We chose an excitation frequency of $\qty{10}{\giga\hertz}$ to exemplify our upper-frequency limit for the experimental verification of our results. The voltage amplitude of $\qty{12.8}{\milli\volt}$ is selected to equal an input power of $\qty[per-mode = symbol]{-25}{\decibel\per\m}$ for a $\qty{50}{\Omega}$ matched setup.

Meshes are tetrahedral and refined so that the maximum element size in current-carrying regions is \(\le 0.3\,\delta_{\rm skin}(\omega_{\max})\), sufficient to resolve skin- and proximity-induced current crowding in CPW conductors~\cite{Mori_2021, ZHANG2018, RAO2019}. Here
\begin{equation}
\delta_{\rm skin}=\sqrt{\frac{2}{\mu_0\,\sigma\,\omega}},
\end{equation}
which gives \(\delta_{\rm skin}^{\rm Au}(\qty{10}{\giga\hertz})\approx \qty{0.8}{\micro\meter}\) for \(\sigma_{\rm Au}\approx 4.1\times10^7~\mathrm{S/m}\). The linear taper is dimensioned (standard CPW synthesis) and verified in FE to remain close to \(\qty{50}{\ohm}\) along its length (low \(|S_{11}|\)). 

\begin{figure*}[th!]
    \centering
    \subfloat[\centering]
    {\includegraphics[height=0.45\columnwidth]{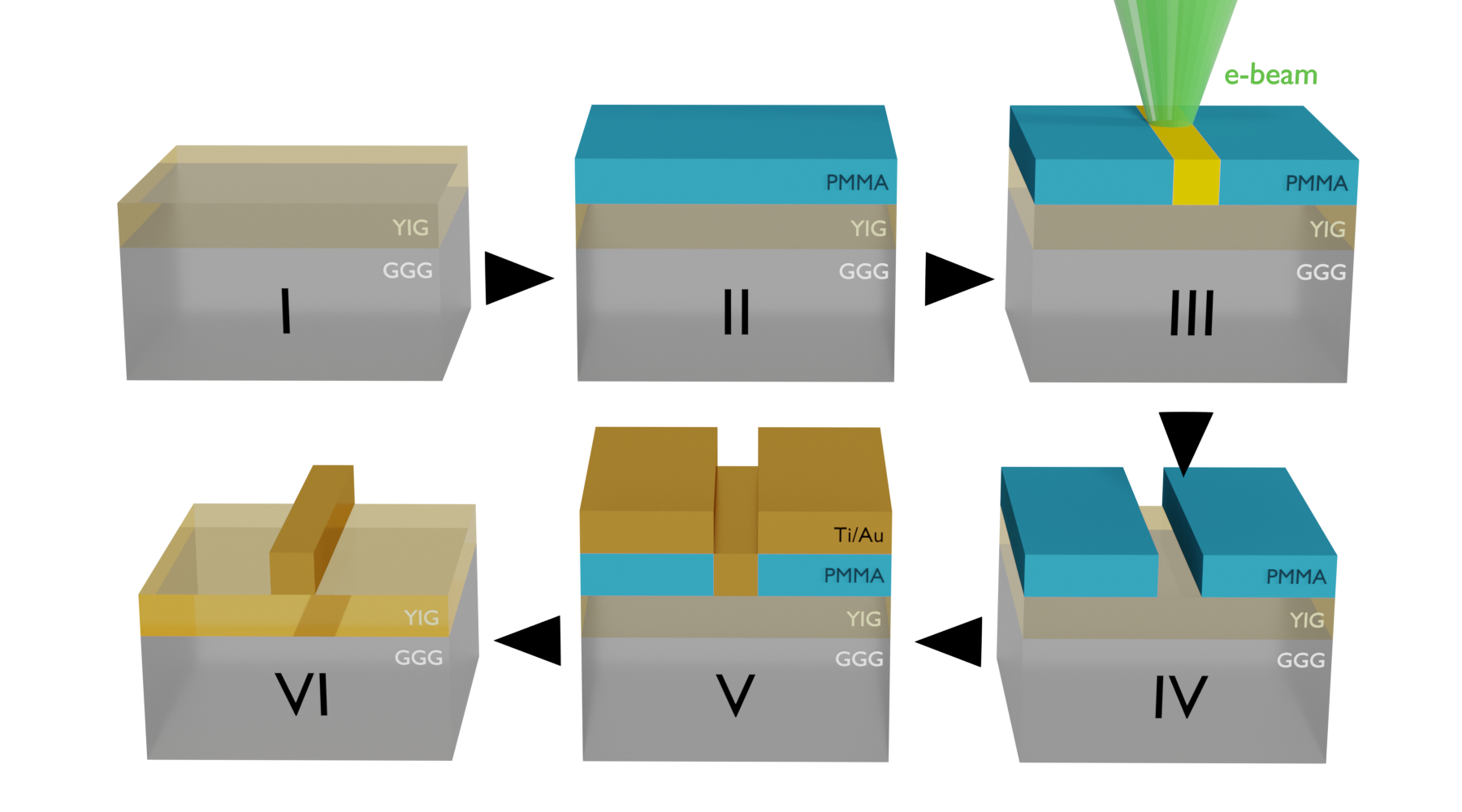} \label{fig:fabrication}}
    \subfloat[\centering]{\includegraphics[height=0.4\columnwidth]{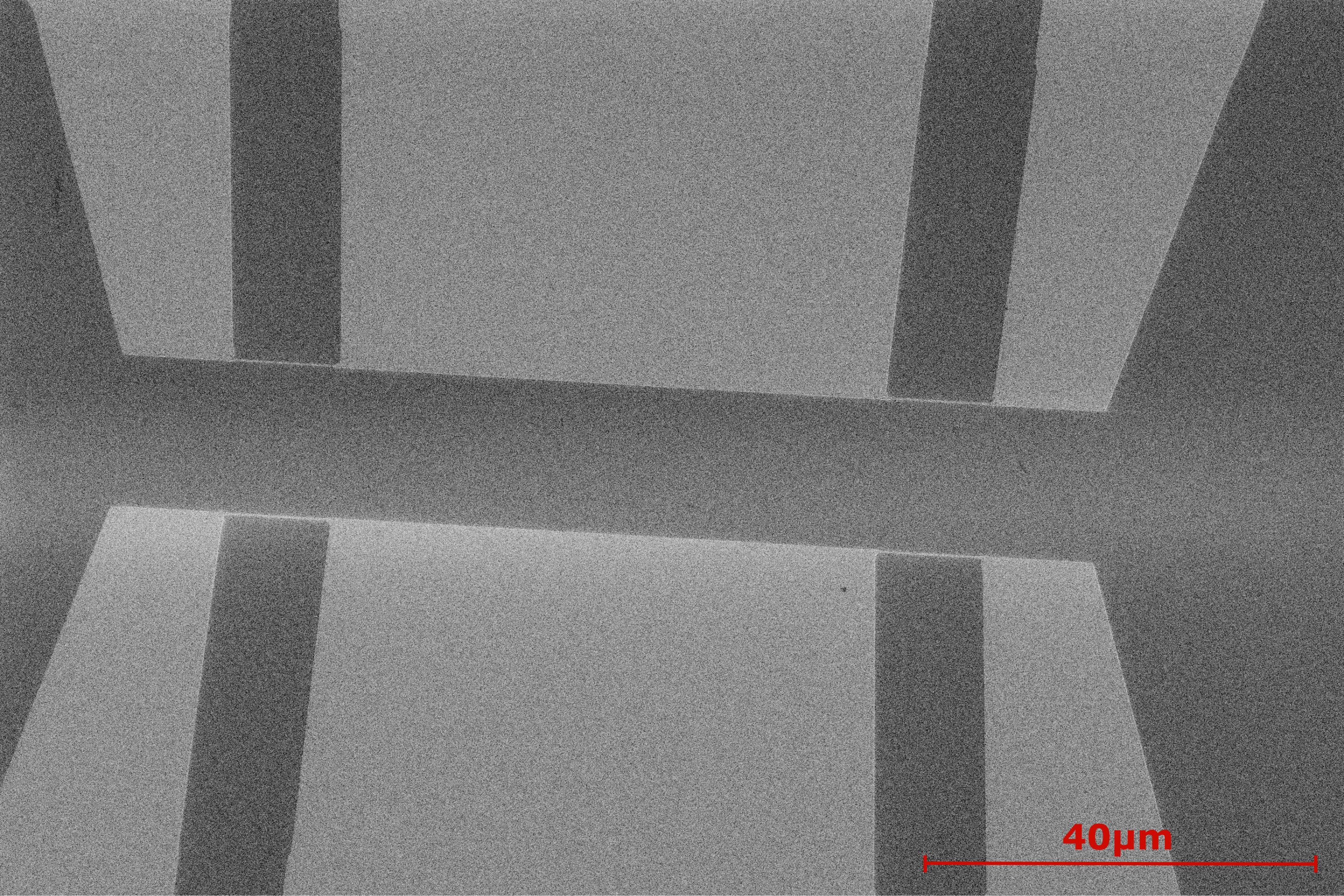} \label{fig:sem_strip}}
    \subfloat[\centering]{\includegraphics[height=0.4\columnwidth]{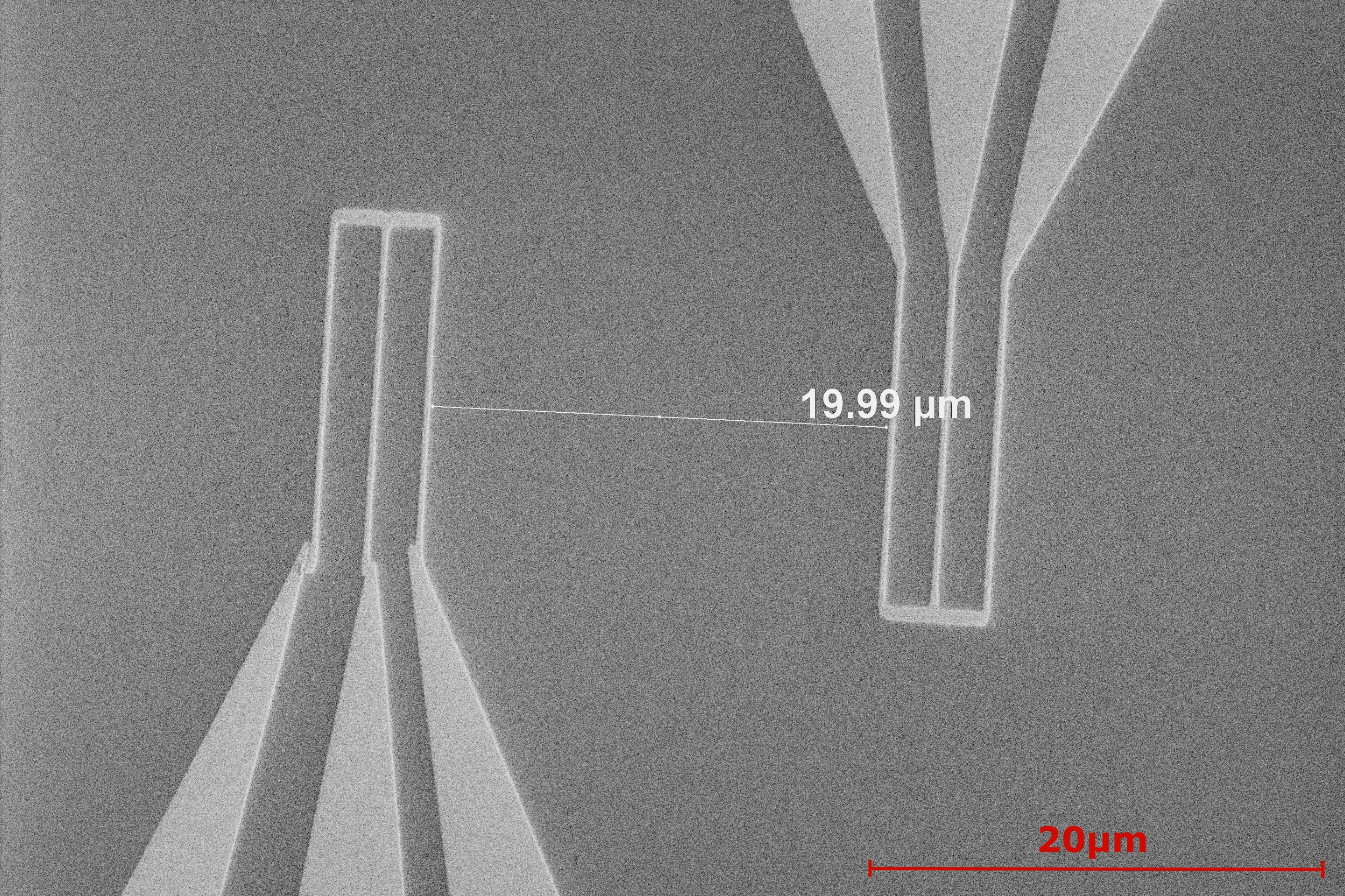} \label{fig:sem_cpw}}
    \hspace{-0.5cm}
    \caption{\textbf{Sample fabrication and results.} (a) Atop a \qty{48}{\nano\m} yttrium-iron-garnet (YIG) film grown on a \qty{500}{\micro \m} thick gadolinium-gallium-garnet (GGG) substrate, the antenna structures are deposited with a nanofabrication process. Using electron-beam lithography and thin-film metal evaporation, the antennas are fabricated. The antennas consit of Ti(\qty{5}{\nano\m})/Au(\qty{85}{\nano\m}). For more information, please refer to the main text. SEM image of (b) the stripline antenna and the CPW tapers, and (c) the CPW antenna with the CPW tapers.}
    \label{fig:fab}
\end{figure*}

\subsection{FE$\to$FD field import and drive component $ \mathbf h_{\!\perp}$}
We export the complex \emph{vector} magnetic-field phasor \(\mathbf h(\mathbf r;\omega)\) on the YIG surface under and around the antenna on a dense Cartesian grid covering the FD box (see Fig.~\ref{fig:simulations}a,b). 
Note, at a timestep at which the current density is at a maximum at the antenna, the magnetic field components are interpolated and exported in a micromagnetic finite-difference (FD) box, indicated in Fig.~\ref{fig:simulations}a,b, underlying the antenna structure. However, the choice of this micromagnetic FD box is arbitrary. It depends on the underlying YIG structure, the specific excitation region being investigated, and, ultimately, the computational power available for the FD micromagnetic simulation.

With \(\hat{\mathbf y}\parallel\mathbf H\), \(\hat{\mathbf x}\parallel\mathbf k\), and \(\hat{\mathbf z}\) out of plane, only the component perpendicular to \(\hat{\mathbf M}_0=\hat{\mathbf y}\) efficiently drives precession,
\begin{equation}
\mathbf h_{\!\perp}(\mathbf r;\omega)=\mathbf h(\mathbf r;\omega)-\big(\mathbf h(\mathbf r;\omega)\cdot\hat{\mathbf y}\big)\,\hat{\mathbf y},
\end{equation}
which preserves the relative phase of \(h_x\) and \(h_z\). We do not take a norm before injection. The FE field is trilinearly interpolated onto the FD mesh. For time-domain drives,
\begin{equation}
\mathbf h_{\!\perp}(\mathbf r,t)=\Re\!\left\{\mathbf h_{\!\perp}(\mathbf r;\omega)\,s_{\rm sinc}(t)\,e^{i\omega t}\right\},
\end{equation}
where \(s_{\rm sinc}(t)\) sets the analysis band, in harmonic sweeps, we inject the stationary phasor directly. 

\subsection{Micromagnetic finite-difference (FD) simulations}

We solve the Landau–Lifshitz–Gilbert equation using \texttt{magnum.np}~\cite{Bruckner2023} with material parameters matching the \qty{48}{\nano\meter} YIG film. The FE drive \(\mathbf h_{\!\perp}(\mathbf r,t)\) is applied as the local RF field. For reference, we use the Kalinikos–Slavin formalism~\cite{Kalinikos_1986} for dispersions and group velocities, and the semi-analytical dispersion relation calculator in \textit{TetraX}~\cite{korberFiniteelementDynamicmatrixApproach2021a}.

\subsection{Figures of merit and spectral definitions}
In \((\mathbf k,\omega)\)-space the driven response obeys 
\begin{equation}
\mathbf m(\mathbf k,\omega)=\boldsymbol\chi(\mathbf k,\omega)\,\mathbf h_{\!\perp}(\mathbf k,\omega),
\end{equation}
the antenna \(\mathbf k\)-weighting is
\begin{equation}
W(\mathbf k,\omega)=\big\|\mathbf h_{\!\perp}(\mathbf k,\omega)\big\|^2,
\end{equation}
and the dispersion-like intensity we plot is
\begin{equation}
I(\mathbf k,\omega)=\big\|\mathbf m(\mathbf k,\omega)\big\|^2,\qquad I(\omega)=\sum_{\mathbf k} I(\mathbf k,\omega).
\end{equation}

\subsection{Why FE\,+\,FD here (and not FDTD)?}

We are in a quasi-static near-field regime (antenna dimensions \(\ll \lambda_{\rm EM}\)) where the physics of interest is the realistic, geometry- and impedance-dependent near-field spectrum delivered to the film, not time-of-flight radiation. A frequency-domain FE model naturally supplies the complex vector \(\mathbf h(\mathbf r;\omega)\) under matched drive and handles tapers, return paths, and multi-material stacks with high fidelity. Other EM solvers (FDTD or FD-EM) could be used in principle. Our choice is for convenience and accuracy in extracting \(\mathbf h(\mathbf r;\omega)\) with the correct impedance context. Another advantage of the presented FE\,+\,FD method here is that no solver is currently available that combines LLG with FDTD.

\section{Fabrication and measurement methods}\label{sec:fabrication}

CPW and stripline antennas are fabricated on the same YIG/GGG chip by e-beam lithography and lift-off of Ti(\qty{5}{\nano\meter})/Au(\qty{85}{\nano\meter}), as shown in Fig.~\ref{fig:fab}a. Secondary-electron images of the realised structures are shown in Fig.~\ref{fig:fab}b,c (stripline/tapers and CPW + tapers). 
First, we spin-coat and bake two layers of positive-tone PMMA on top of the YIG sample (step I) with a total thickness of \qty{230}{\nano\m}, covered by a layer of conductive material (step II). The structure is patterned into the resist using electron-beam lithography (step III), followed by a resist development (step IV). After deposition of Ti(\qty{5}{\nano\m}/Au(\qty{85}{\nano\m} using electron-beam physical vapour deposition the unexposed regions are lifted-off (step V-VI).
Due to fabrication tolerances, the stripline width is \((225\pm19)\,\mathrm{nm}\) over \(\qty{15}{\micro\meter}\) length; the CPW has \(\qty{15}{\micro\meter}\) length with \((275\pm15)\,\mathrm{nm}\) signal/ground widths and \(\qty{20}{\micro\meter}\) inner-ground spacing. The CPW transmission lines feeding the tapers are \(\qty{2}{\milli\meter}\) long. The tapers are designed linearly between the \(\qty{50}{\ohm}\) impedance-matched CPW transmission lines and CPW/stripline antennas, using a mode solver.

We measure the \(S\)-parameters $\Delta S(\omega)_{\mathrm{21}}$ as a function of bias field and extract both the amplitude and phase. $\Delta S(\omega)_{\mathrm{21}}= S(\omega)_{\mathrm{21}} - S (\omega)_{\mathrm{21,ref}}$ denotes the background corrected signal. 

Microwave measurements use GGB 40A probes and an Anritsu MS4642B VNA; the sample sits in a GMW 3473-70 electromagnet (air gap \(\qty{8}{\centi\meter}\), pole diameter \(\qty{15}{\centi\meter}\)), providing up to \(\qty{900}{\milli\tesla}\). Unless stated otherwise, the applied RF power is \(-25~\mathrm{dBm}\), to maintain the linear excitation regime. The bias field is applied along \(\hat{\mathbf y}\) as defined above. 

\section{Results and validation}\label{sec:results}
\subsection{Antenna near-fields and \texorpdfstring{\(k\)}{k}-weighting}

\begin{figure}[tbh!]%
    \subfloat[\centering]
    {{\includegraphics[width=0.9\columnwidth]{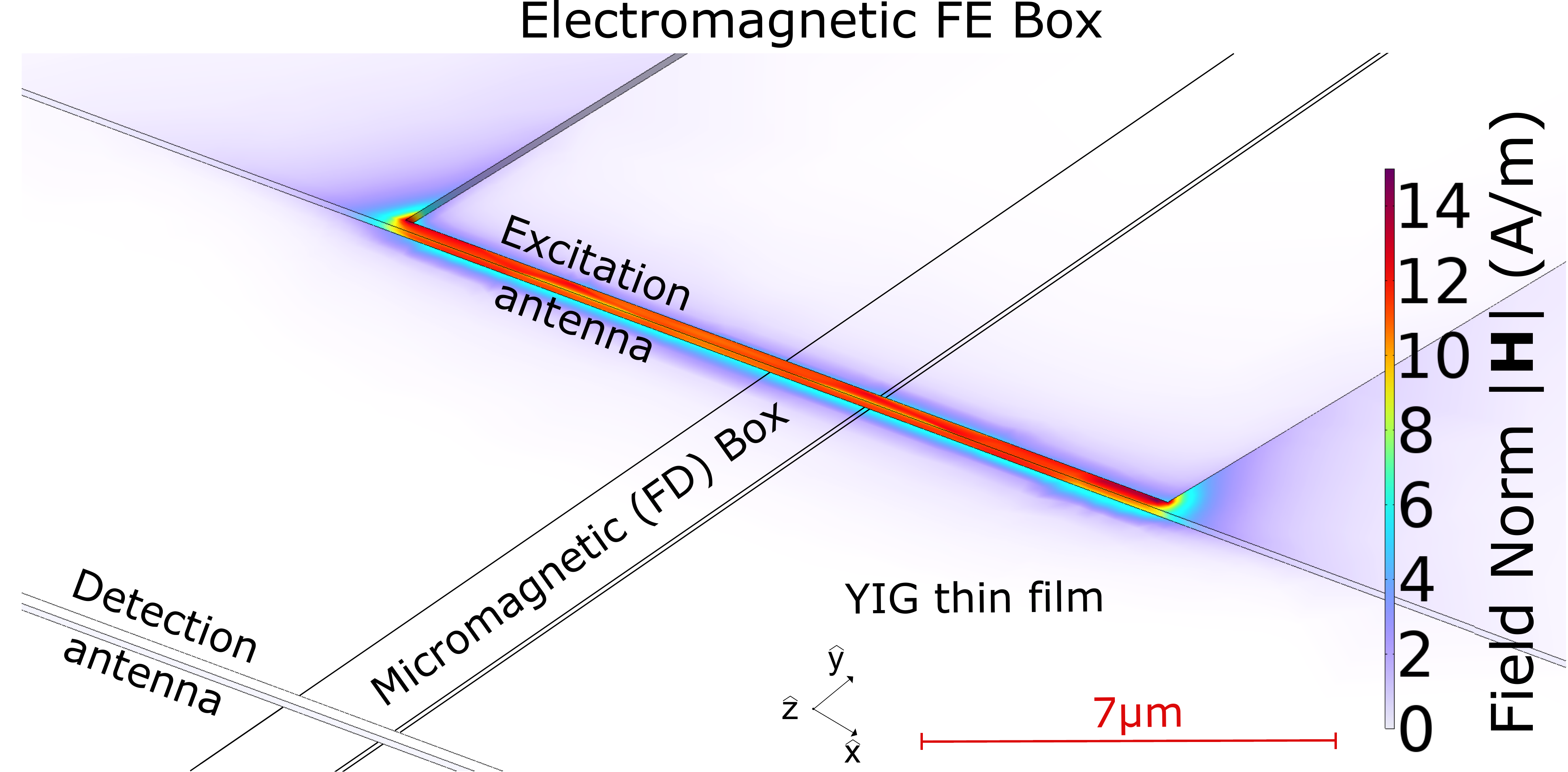}\label{fig:strip_sim}}}
    \hspace{-1.5cm}
    \subfloat[\centering]{\includegraphics[width=0.9\columnwidth]{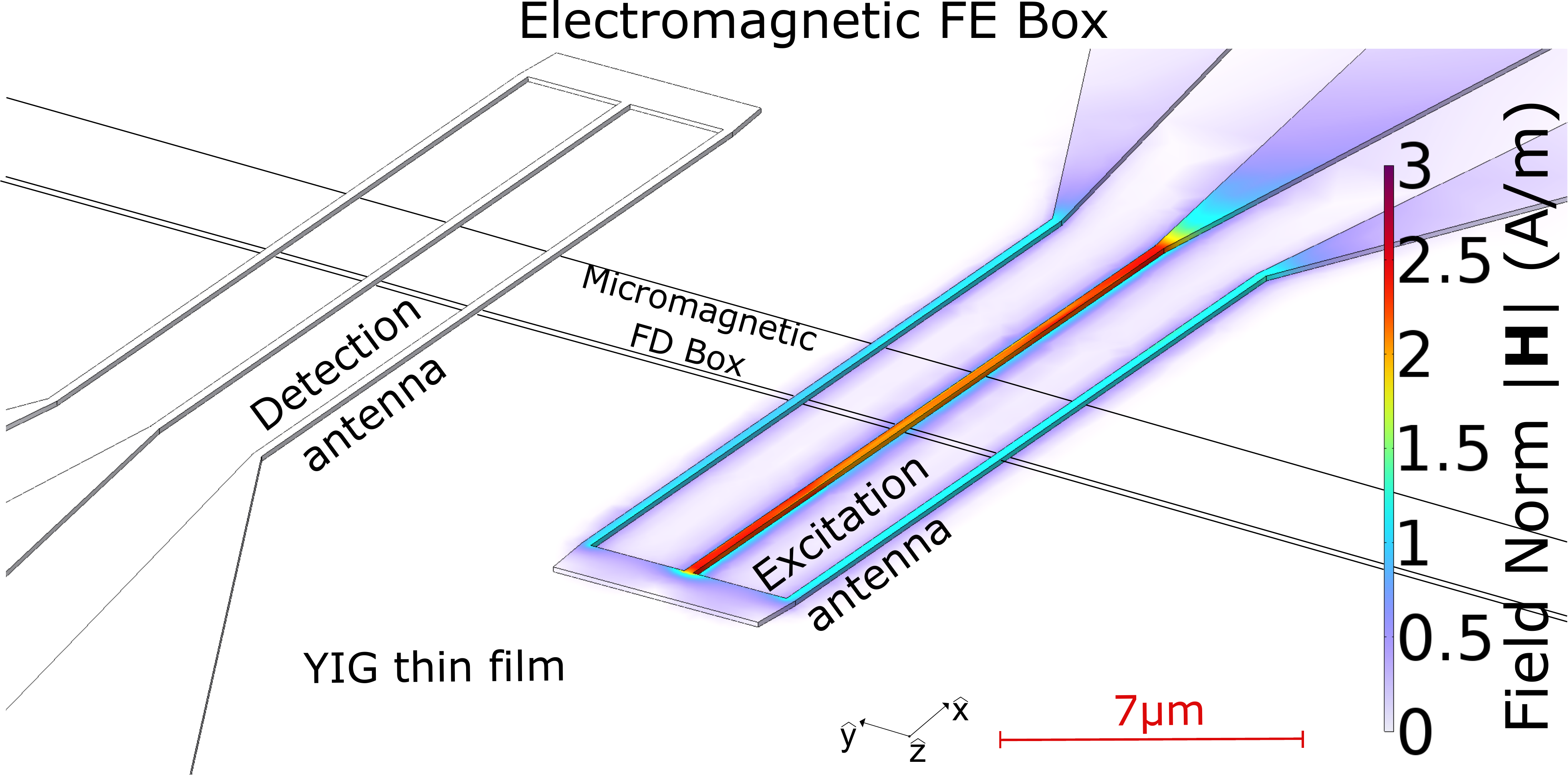} \label{fig:cpw_sim}} 
    \hspace{-1.5cm}
    \subfloat[\centering]{\includegraphics[width=0.45\columnwidth]{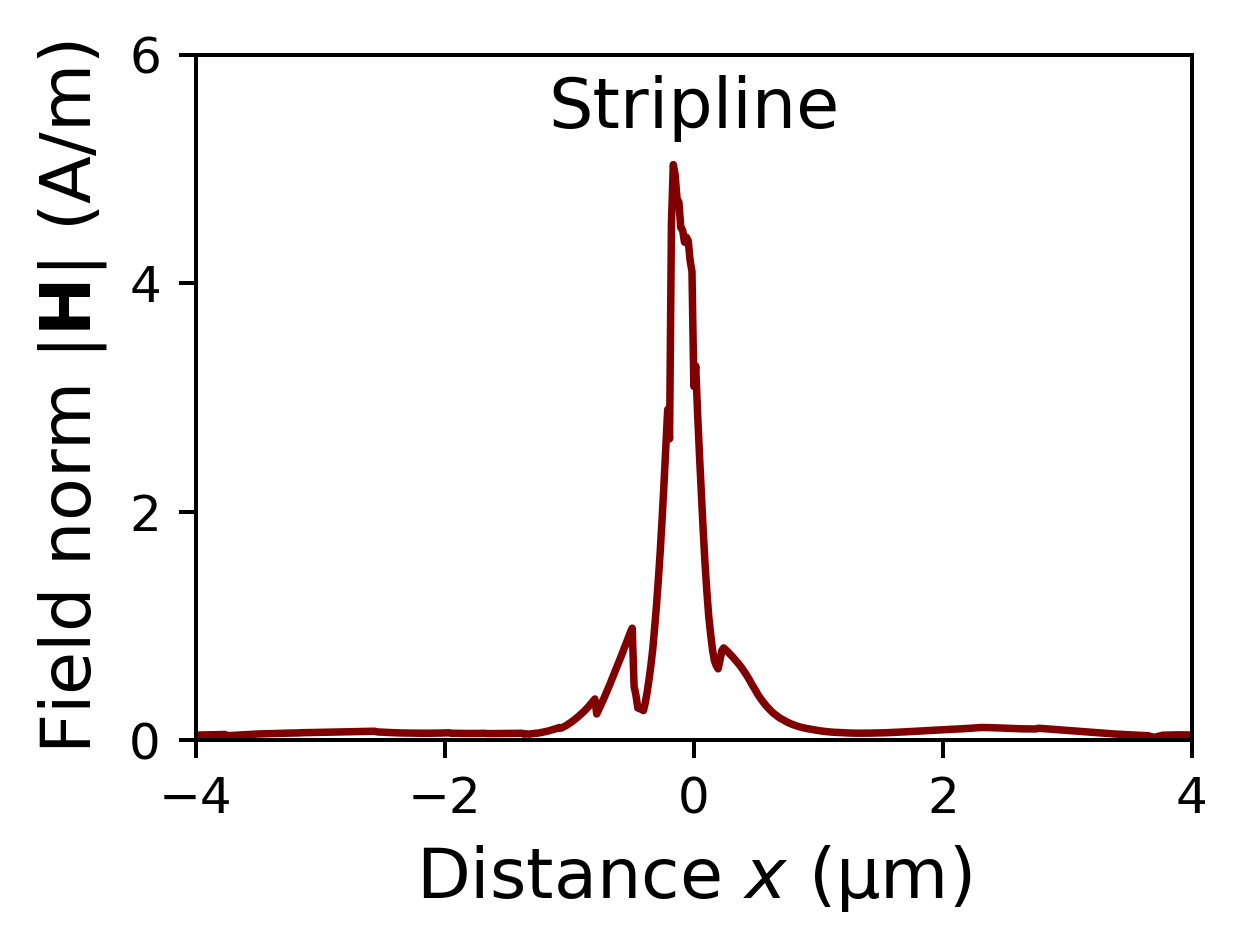} \label{fig:strip_hfield}}
    \subfloat[\centering]{\includegraphics[width=0.45\columnwidth]{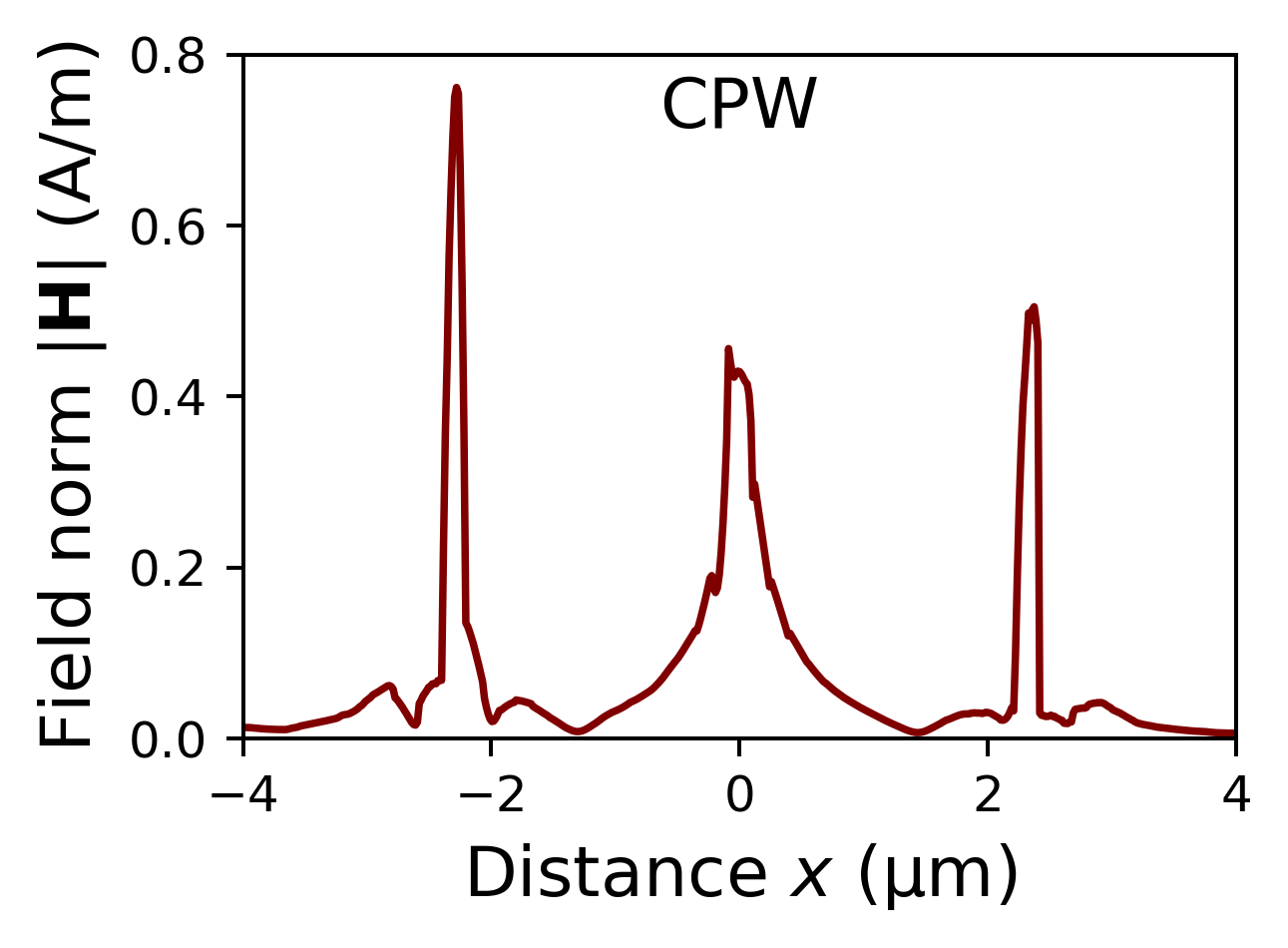} \label{fig:cpw_hfield}}
    \hspace{-1.5cm}
    \subfloat[\centering]{\includegraphics[width=0.45\columnwidth]{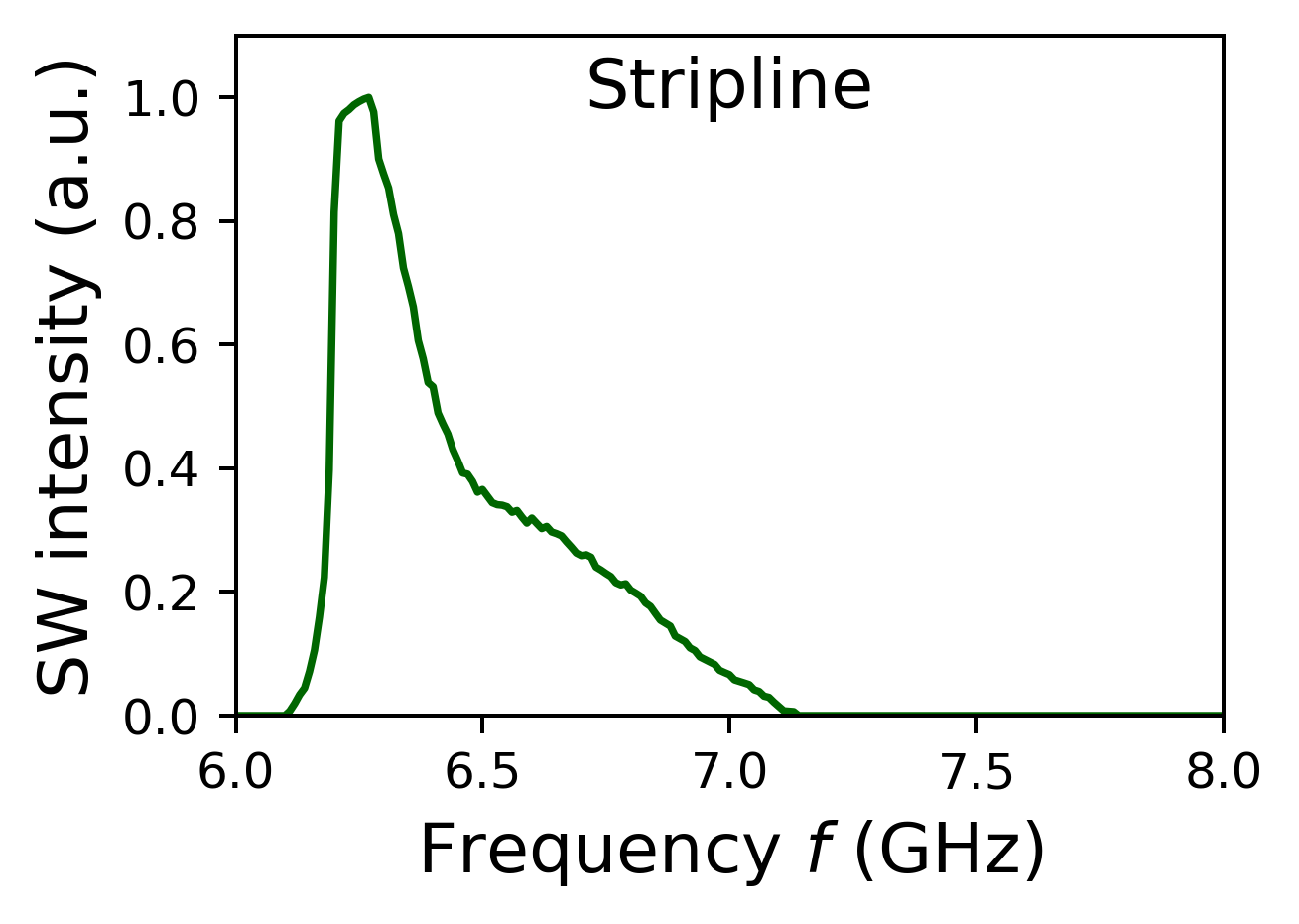} \label{fig:strip_exceff}}
    \subfloat[\centering]{\includegraphics[width=0.45\columnwidth]{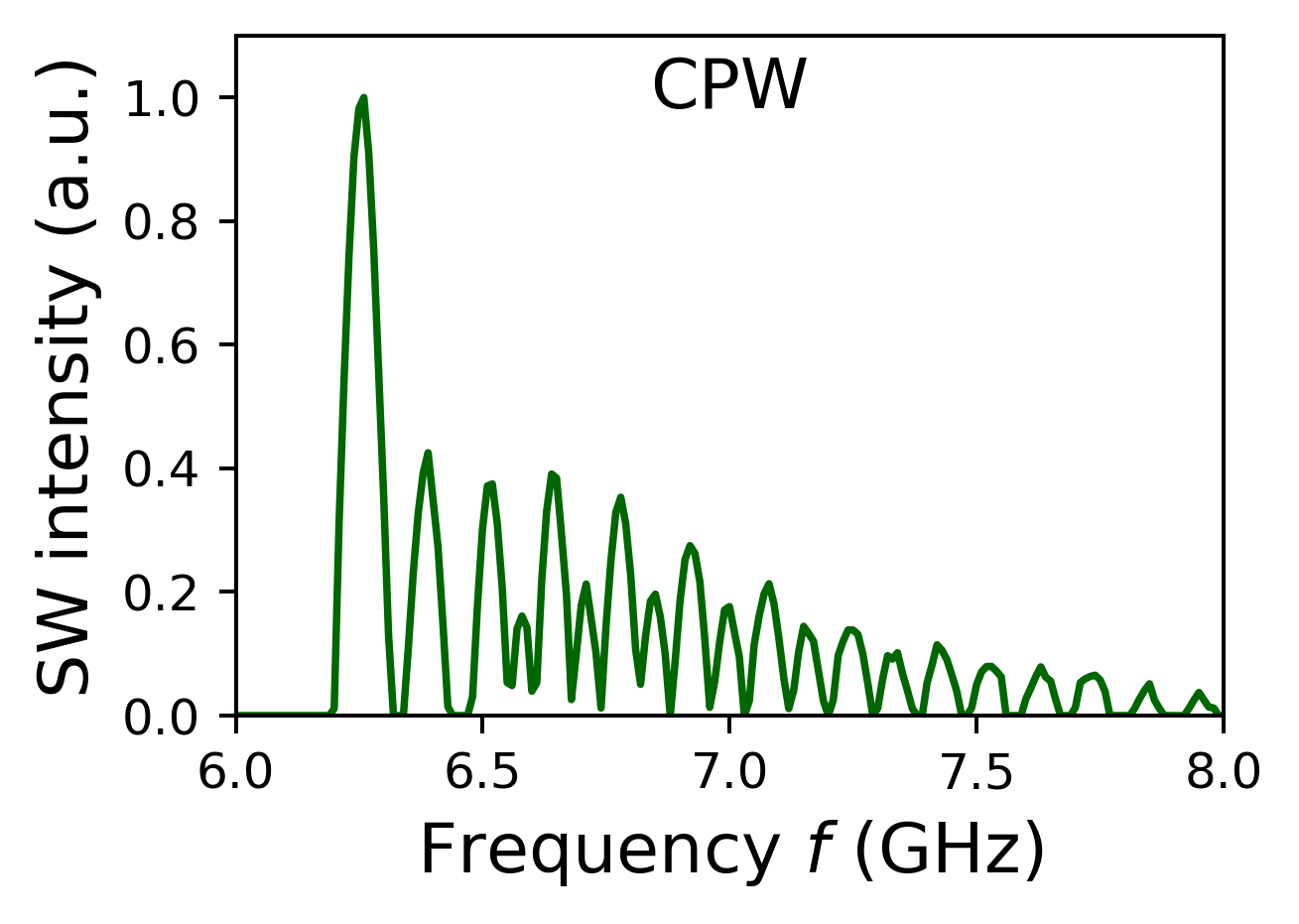} \label{fig:cpw_exceff}}
    \vspace{0.3cm}
    \caption{\textbf{FE and FD simulation results.} Magnetic field norm on the surface of the simulated (a) stripline antenna and (b) coplanar waveguide antenna in A/m represented by the false colour bar. The components of the magnetic field from the excitation antenna are exported in the region of the micromagnetic finite-difference (FD) box. Averaged over the width of the FD box, the magnetic field norm is plotted for the (c) stripline antenna and (d) CPW antenna 
    . In the FD box, micromagnetic simulations are conducted using the excitation field gained from the electromagnetic simulations. The normalised spin-wave excitation intensities are extracted from the micromagnetic simulations for (e) the stripline antenna and (f) the CPW antenna. 
    }%
    \label{fig:simulations}
\end{figure}

Figure~\ref{fig:simulations}a,b show colour-coded maps of the magnetic field at the YIG surface under stripline and CPW exciters. Linecuts averaged over the FD-box width (Fig.~\ref{fig:simulations}c,d) reveal a three-peak CPW profile: a central maximum from the signal line and two side maxima from the ground-return currents. With approximately equal return splits, each ground carries \(\sim I/2\), yielding about half-amplitude side peaks, the narrower apparent width of the ground-line peaks follows from edge current crowding due to skin/proximity effects~\cite{Mori_2021, ZHANG2018, RAO2019}. Fourier analysis gives the antenna \(k\)-weighting \(W(k,\omega)\), where strip-like exciters show a main selection scaling approximately as \(k_{\rm peak}\approx \pi/w\), modified by the current return geometry and the finite distance to the film. 

\subsection{Simulated dispersion and group velocity}
\begin{figure*}[th!]%
    \centering
    \subfloat[\centering]
        {\includegraphics[height=0.6\columnwidth]{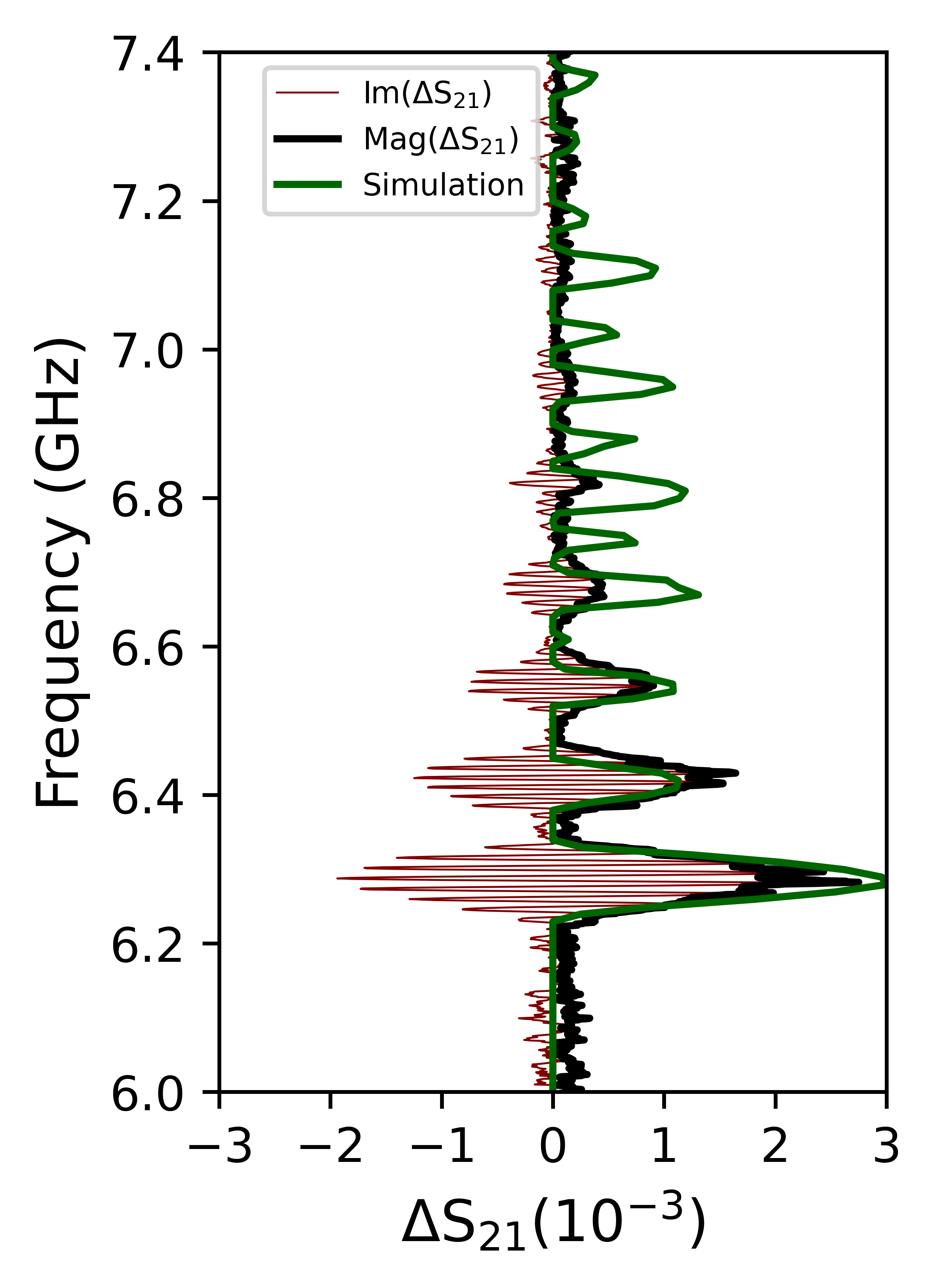} \label{fig:CPW_AEPSWS}}
    \subfloat[\centering]{\includegraphics[height=0.6\columnwidth]{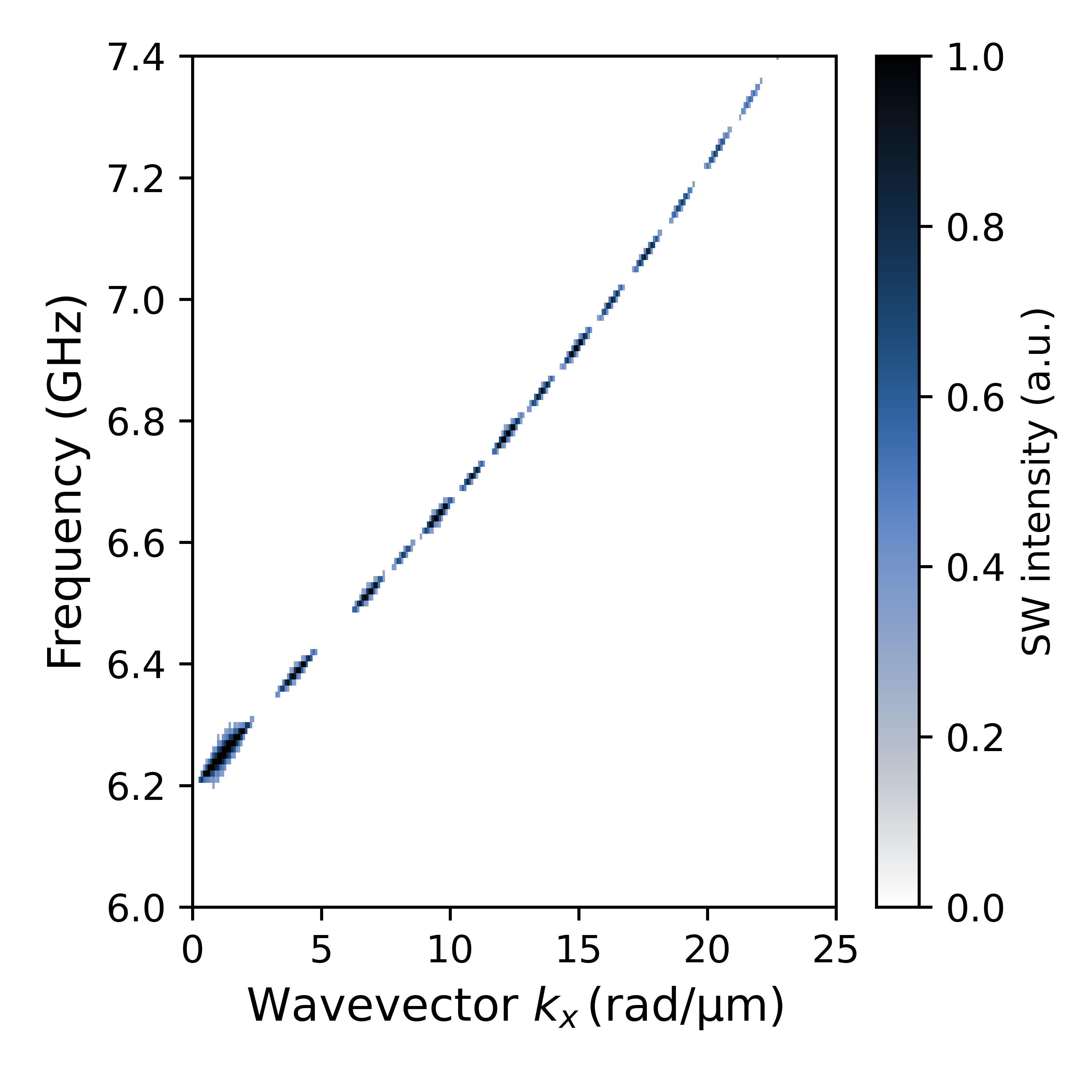} \label{fig:dispersion_cpw}}
    \subfloat[\centering]{\includegraphics[height=0.6\columnwidth]{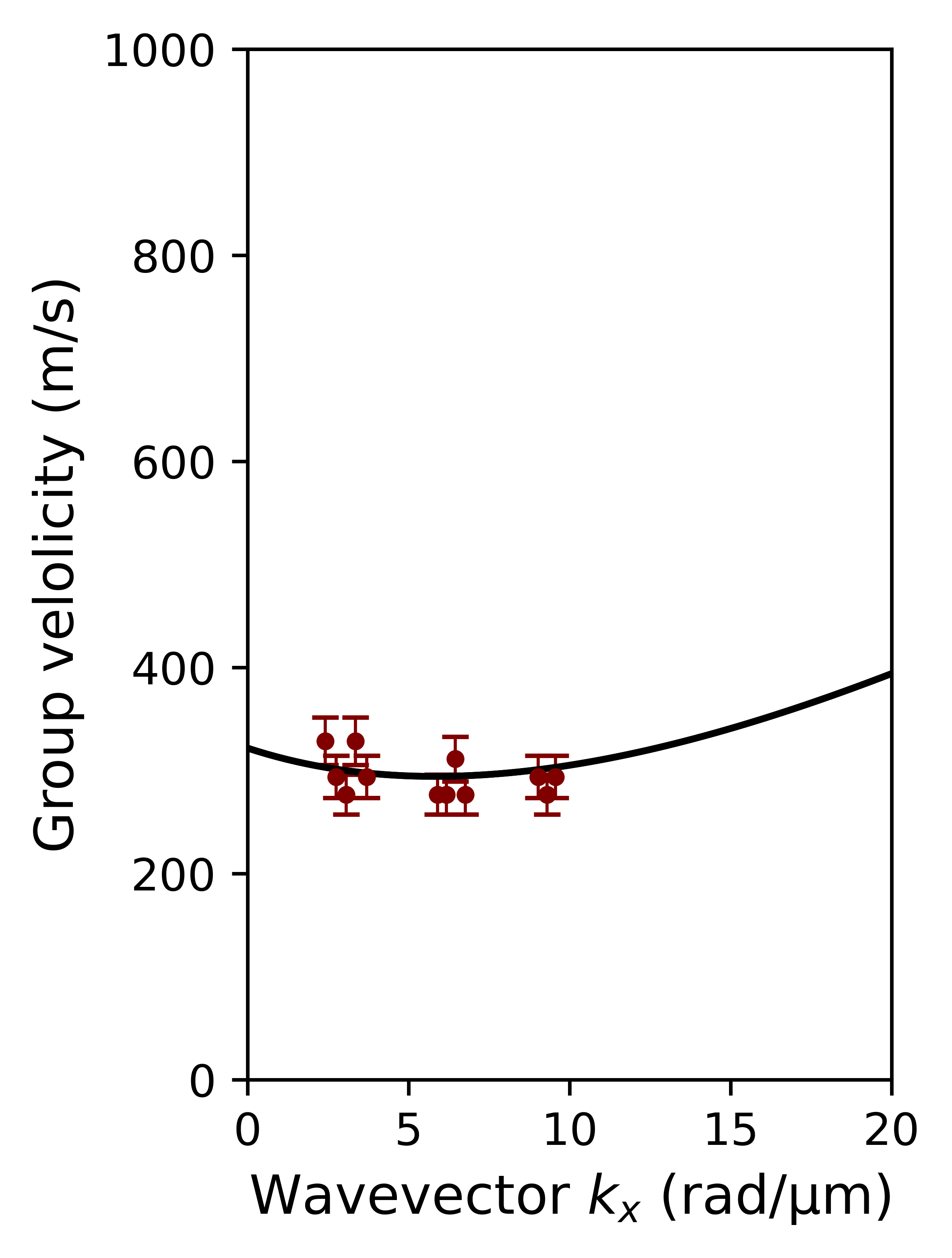} \label{fig:group_vel_cpw}}
    \hspace{-0.5cm}
    \subfloat[\centering]
        {\includegraphics[height=0.6\columnwidth]{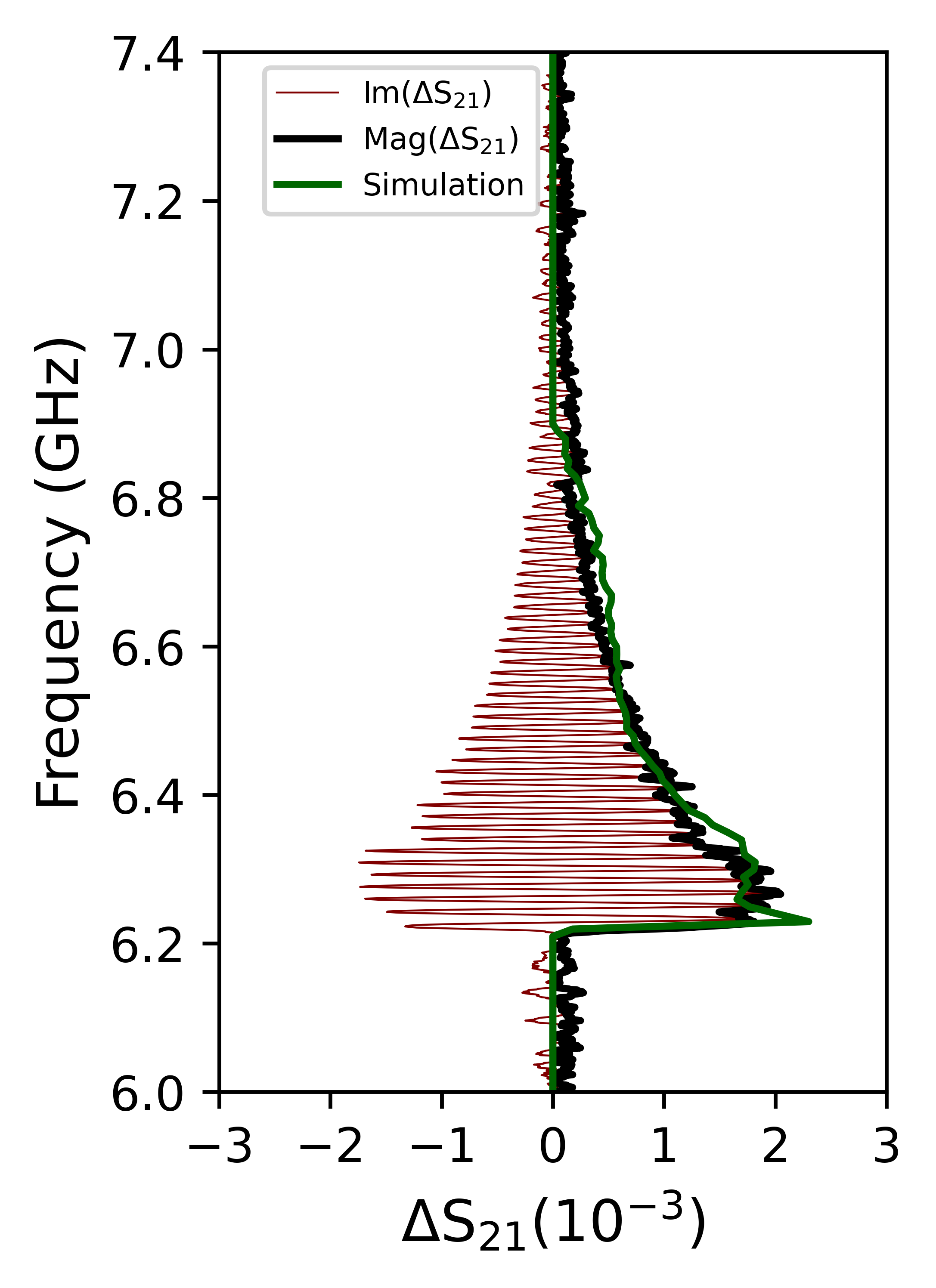} \label{fig:strip_AEPSWS}}
    \subfloat[\centering]{\includegraphics[height=0.6\columnwidth]{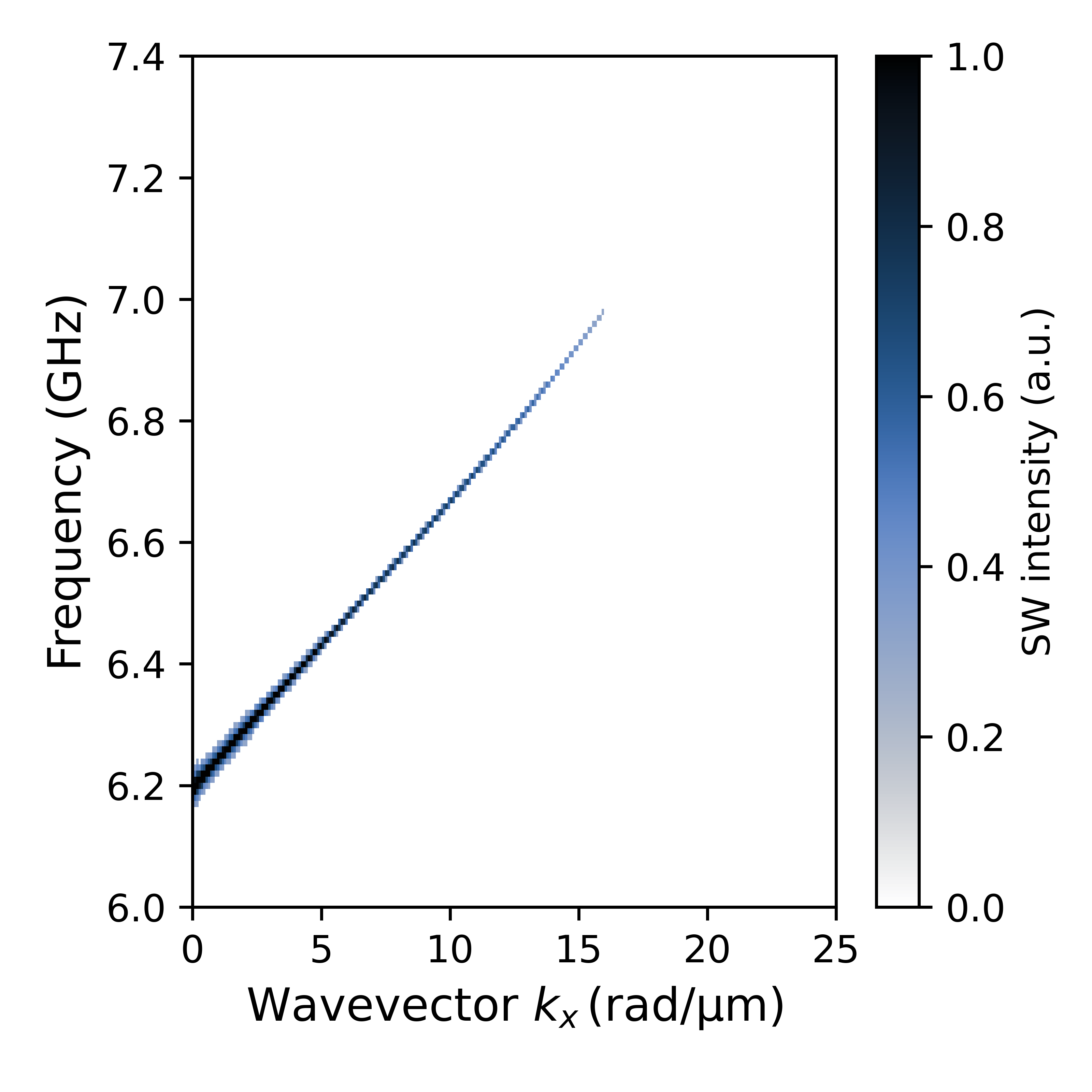} \label{fig:dispersion_strip}}
    \subfloat[\centering]{\includegraphics[height=0.6\columnwidth]{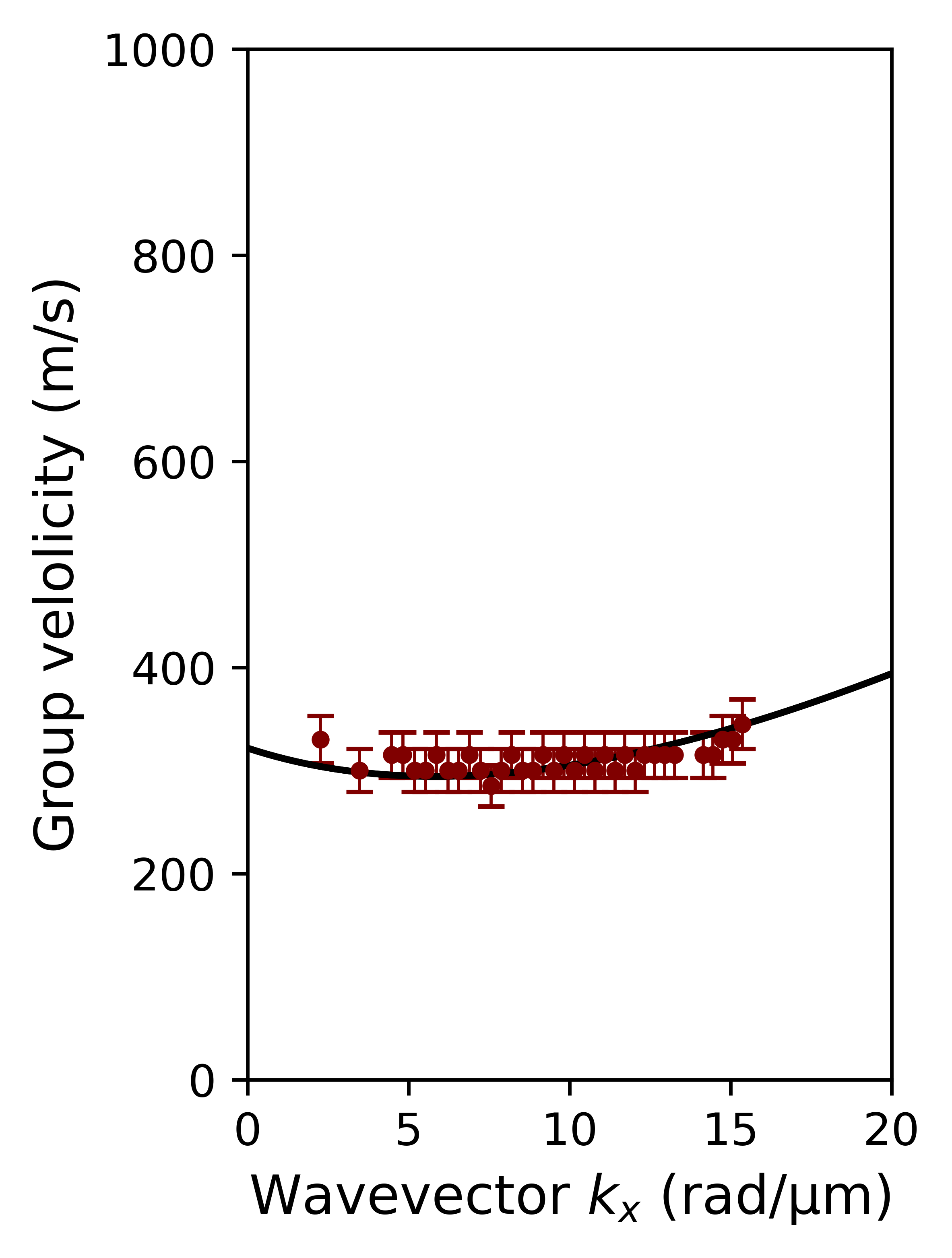} \label{fig:group_vel_strip}}
    \caption{\textbf{All-electrical propagating spin-wave spectroscopy results compared with the simulation.} 
    The linear magnitude and imaginary part of the $\Delta S_{21}$ parameter compared to the simulated spin-wave excitation in
    Demon-Eshbach mode at an external magnetic field of \qty{150}{\milli \tesla} for (a) the CPW antenna and (d) the stripline antenna are shown. The dispersion relation was obtained by the numerical simulation using the excitation fields for (b) the CPW antenna and (e) the stripline antenna, illustrated by the colour map. Experimentally obtained group velocities for (c) the CPW antenna and (f) the stripline antenna, compared with the theoretically obtained group velocities by the derivative of the dispersion relation obtained by the \textit{Kalinikos-Slavin} model~\cite{Kalinikos_1986} (solid black line).}%
    \label{fig:experiment}
\end{figure*}

Using the Kalinikos-Slavin model and \textit{TetraX} we compute \(\omega(k)\) and \(v_g(k)=\partial\omega/\partial k\) for the \qty{48}{\nano\meter} YIG film. The driven intensity \(I(k,\omega)\) forms ridges along \(\omega(k)\), broadened by the antenna 
\(k\)-weighting \(W(k,\omega)\). Summation over \(k\) yields \(I(\omega)\), used below for quantitative comparison with \(\Delta S_{21}\).

\subsection{Quantitative comparison with AEPSWS}
For both antennas we measure \(\Delta S_{21}\) and use the oscillation period in \(\operatorname{Im}\,\Delta S_{21}\) to extract the experimental group velocity via the standard PSWS relation
\begin{equation}
v_g^{\rm exp}(\omega)=\Delta f(\omega)\,D,
\label{eq:vg_exp}
\end{equation}
with \(D\) the antenna spacing and \(\Delta f\) the spacing of adjacent extrema~\cite{Vlaminck2010,Knauer2023}. From the simulation, 
\begin{equation}
v_g^{\rm th}(\omega)=\partial\omega/\partial k
\label{eq:vg_th}
\end{equation}
is taken along the ridge of \(I(k,\omega)\). 
We report \(\mathrm{RMSE}[f_{\rm ridge}(\omega)-f_{\rm peak}(\omega)]\) and \(\mathrm{RMSE}[v_g^{\rm th}-v_g^{\rm exp}]\).

At \(H=\qty{150}{\milli\tesla}\), Fig.~\ref{fig:experiment} compares \(I(\omega)\) with the measured linear magnitude and imaginary part of \(\Delta S_{21}\), and displays \(I(k,\omega)\) alongside the calculated dispersions.
The data points are connected for clarity. Furthermore, the intensities of spin wave excitations obtained from the numerical simulations (see Fig.~\ref{fig:simulations}, green line) are included for comparison and validation of the numerical results.
The simulated excitation intensities are presented in arbitrary units in the FD simulation. These intensities are scaled by an arbitrary factor to align with the experimental spin-wave amplitude, ensuring better comparability between simulation and measurement results.
The FMR frequency ($k=0$) and the spin wave transmission spectrum between measurements and the simulation agree well for both RF antennas.
In the case of the CPW antennas (Fig~\ref{fig:CPW_AEPSWS}), we note that the measurements are unable to fully capture the spin-wave magnitude at higher wavevectors, as predicted by the simulation. This discrepancy may be attributed to the fabrication accuracy of the antennas. We obtained a spin-wave signal transmission ratio of up to $\qty{0.6}{\percent}$ for the CPW antenna and $\qty{0.4}{\percent}$ for the stripline antenna over a spin-wave transmission distance of $\qty{20}{\micro\m}$, when comparing the analytical with the measured results.

To further compare the experimental results with the simulations, we present the spin wave intensity obtained from the micromagnetic FE simulations for the CPW antenna shown in Fig.~\ref{fig:dispersion_cpw} and the stripline antenna depicted in Fig.~\ref{fig:dispersion_strip}. The intensities, as indicated by the colour bar, are plotted against frequency and the corresponding wavevector. These simulated intensities represent the dispersion relations. We find that the dispersion curves align well with the measured spin-wave signals. 
Figures~\ref{fig:group_vel_cpw} and~\ref{fig:group_vel_strip} show the group velocities as the derivative of the group velocity equation~\ref{eq:vg_th} indicated by the black solid line. The experimental values for the group velocities are given by equation~\ref{eq:vg_exp} and indicated by the red dots, with $\Delta f$ the periodicity of the oscillations in $\textrm{Im}(\Delta S_{21})$ and $D$ the gap between the antennas~\cite{Vlaminck2010}.
For better comparability with literature~\cite{Vlaminck2010, Knauer2023} we plot the group velocity in a range reaching from $\qty{0}{\meter\per\second}$ to $\qty{1000}{\meter\per\second}$. The error bars for the measured group velocities are limited by the frequency resolution of the measured $S_{21}$-parameter. We observe that the measured group velocities agree well with the theoretically obtained ones.
Our data from Figs.~\ref{fig:simulations} and~\ref{fig:experiment} can be summarised in the values in the following Table~\ref{table:summary}. Note, for our plots and calculation we keep the antenna spacing as a free fit parameter and find \(D^{\mathrm{fit}}_{\mathrm{CPW}}=22.7\, \mathrm{\mu m}\) and \(D^{\mathrm{fit}}_{\mathrm{stripline}}=19.7\, \mathrm{\mu m}\). Both values are well within the fabrication and measurement tolerances. Additionally, the field in the CPW is more widely spread (see Fig.~\ref{fig:simulations}).

\begin{table}[ht] 
\caption{Comparison of experimental the theoretical values found in Figs.~\ref{fig:simulations} and~\ref{fig:experiment}. } 
\centering  
\begin{tabular}{c c c} 
\hline\hline 
 & CPW &  Stripline \\ [0.5ex]
\hline 
{\(\mathrm{RMSE}[v_g^{\rm th}-v_g^{\rm exp}]\)} (\si{\meter\per\second})& 20.6 &  10.1 \\ 
{\(\mathrm{Rel.~RMSE}[v_g^{\rm th}-v_g^{\rm exp}]\)} (\%)  & 6.8 & 3.4 \\ 
\(\mathrm{RMSE}[f_{\rm ridge}(\omega)-f_{\rm peak}(\omega)]\) (MHz)& 35.1 & 10.6\\ 
\(\mathrm{Rel.~RMSE}[f_{\rm ridge}(\omega)-f_{\rm peak}(\omega)]\) (\%)& 0.5  & 0.2 \\ [1ex]
\hline 
\end{tabular}
\label{table:summary}
\end{table}

\begin{figure}[t!]%
    {\includegraphics[height=0.8\columnwidth]{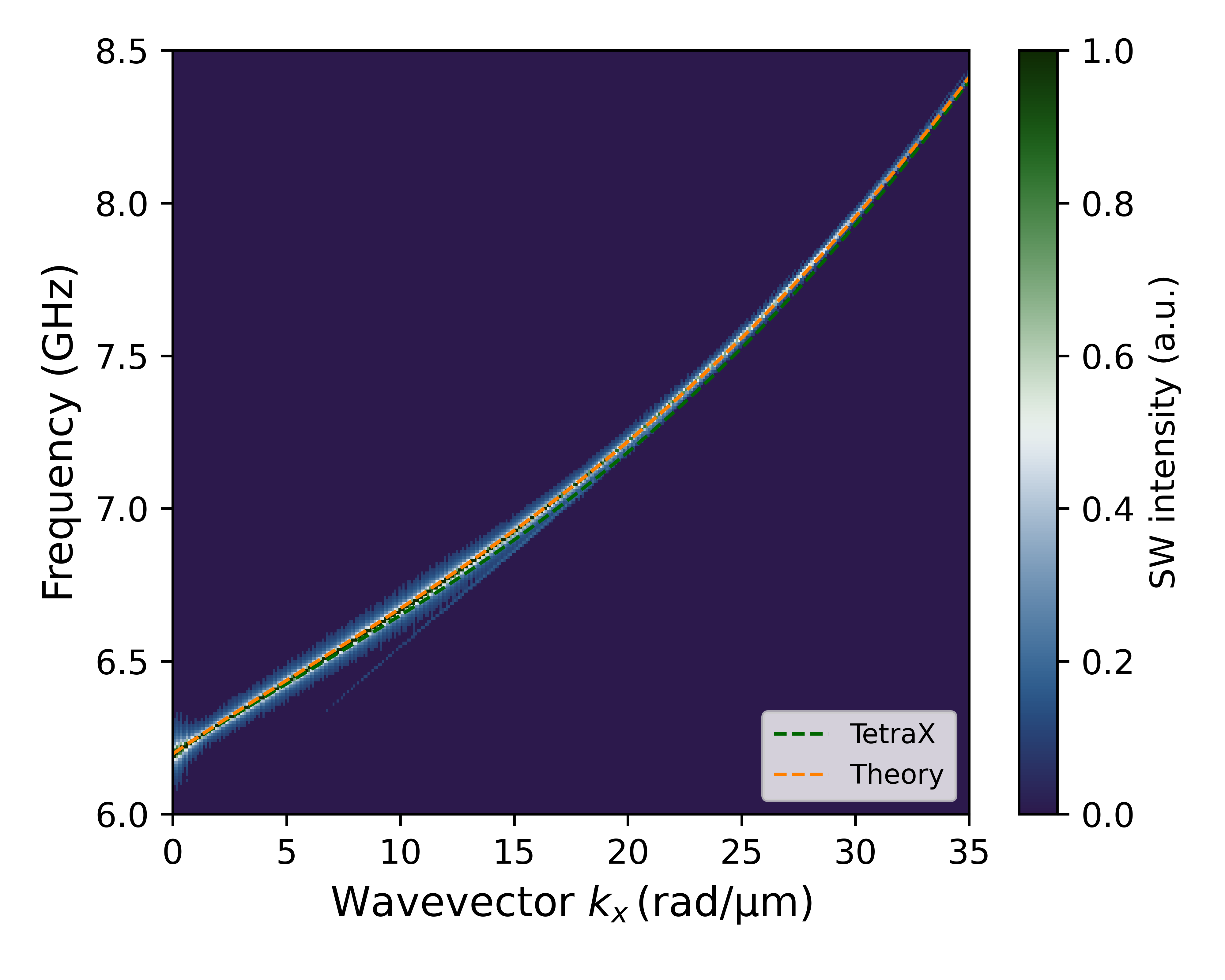}}
    \caption{\textbf{Comparison between numerical simulation, \textit{TetraX} calculations and calculations obtained by the \textit{Kalinikos-Slavin} model.} The Hue of the colourmap represents the dispersion relation obtained by the numerical simulations. Dispersion relations obtained by \textit{TetraX} calculations and \textit{Kalinikos-Slavin} model are represented by the dashed green and orange lines, respectively.}%
    \label{fig:dispersion}
\end{figure}

To further underpin the simulated dispersion results from Figs.~\ref{fig:dispersion_cpw} and~\ref{fig:dispersion_strip}, we compare them with numerical calculations obtained by the \textit{TetraX} and theoretical calculations, as shown in Figure~\ref{fig:dispersion}. The dispersion relations obtained theoretically are calculated using the \textit{Kalinikos-Slavin} model. The figure illustrates the three results for the CWP antenna: the numerical simulations are represented by the false contour plot, the black dashed line indicates the numerical calculations, and the analytical calculations are shown with the red dashed line. The three results are in good agreement, demonstrating that the dispersion obtained from the numerical finite difference simulations, derived by coupling electromagnetic finite element simulations, is verifiable.


\subsection{Design rules: CPW vs.\ stripline; tapers and returns}

The FE\(\to\)FD chain suggests practical rules. (i) Reducing the signal-line width increases \(k_{\rm peak}\approx \pi/w\), enabling access to higher \(k\) at the cost of reduced overlap with long-wavelength susceptibility. (ii) Ground-return symmetry sets the side-lobe balance in CPW \(W(k,\omega)\) and intentional asymmetry suppresses one side-lobe and reduces parasitic counter-propagating content. (iii) The taper/return geometry controls the \(h_x/h_z\) mix and edge current crowding near inner CPW edges enhances this component mixing~\cite{Mori_2021, ZHANG2018, RAO2019}. These levers are visible directly in the component-resolved near-field maps and their Fourier spectra.

A further expansion of these design rules for future research includes the calculation of spin-wave resistance and excitation efficiency, which can be obtained from micromagnetic simulations by analysing the dynamic magnetisation response to an applied excitation field and calculating the induced voltage in the transducer. This expansion enables the quantification of how efficiently electrical energy is converted into propagating spin waves and how much energy is dissipated, as discussed in, for example,~\cite{Erdelyi2025, Bruckner2025, Kohl2025}. 

Coupling micromagnetic simulations with classical electromagnetic solvers, such as \textit{COMSOL}, allows for a more accurate description of effects like the skin effect and proximity effect in the metallic transducer. Meanwhile, the interaction with the YIG film is treated micromagnetically, assuming no significant deviation from the provided excitation field. Such hybrid approaches enable realistic modelling and systematic optimisation of transducer geometries for minimal loss and maximal coupling efficiency.

\section{Conclusions and outlook}\label{sec:conclusion}

We established a vector-faithful workflow that couples a realistic field-distributed, impedance-matched nanoantenna to propagating spin waves in YIG by importing the full complex near-field into micromagnetics. By separating antenna filtering \(W(k,\omega)\) from magnetic response via \(\boldsymbol\chi\), we obtain quantitatively \(W\). The CPW’s characteristic three-peak drive profile, the \(\sim\pi/w\) \(k\)-selection of strip-like exciters, and the role of return-path geometry emerge naturally and are validated against AEPSWS on the same YIG/GGG stack (Figs.~3–5). The FE stage already incorporates impedance, skin effect, and multi-material details, while the FD stage can sweep geometry to target specific \(k\)-bands. With these levers, agreement with the experiment follows from the physics rather than being the endpoint.

Specifically, the ability to include multi-material details will enable future work to incorporate temperature-dependent conductivities or other loss channels, which are crucial for advancing magnon-based quantum technologies, especially on-chip magnonic transducers.

\subsection*{Acknowledgements}
This project has received funding from the European Union’s Horizon 2020 research and innovation programme under the Marie Sklodowska-Curie grant agreement No.~101025758, project OMNI. 
The authors acknowledge CzechNanoLab Research Infrastructure supported by MEYS CR (LM2018110).
The work of ML was supported by the German Bundesministerium für Wirtschaft und Energie (BMWi) under Grant No.~49MF180119. 
CD thanks R. Meyer (INNOVENT e.V.) for technical support.
This research was funded in part by the Austrian Science Fund (FWF) through the [10.55776/I6568] project Paramagnonics, the [10.55776/PAT3864023] project IMECS, the [10.55776/PIN1434524] project MagNeuro, and [10.55776/I6068].

\subsection*{Author Contributions}
AH fabricated the samples and performed the experiments with support from DS and under the guidance of SK. AH analysed and interpreted the data with support from AVC and SK. AH performed the electromagnetic simulation. AH performed the micromagnetic simulation supported by AAV, SaK, FB, and CA under the support of DiS. CD, ML and TR prepared the LPE sample. CD conceived and supervised the growth of LPE film. SK conceived the project idea and led the project. All
authors discussed the results.
AH and SK wrote the manuscript with support from all co-authors. 

\subsection*{Conflict of interest} 
The authors declare no conflict of interest.



\subsection*{Correspondence and requests for materials}
The data supporting this study's findings are available from doi....

\bibliography{ref.bib}



\end{document}